\documentclass[preprint,12pt]{elsarticle}
\usepackage{graphicx,amsmath,amssymb,amstext}
\usepackage{amssymb,amsbsy,amsfonts,amsthm,color}
\usepackage{mathtools,nccmath}

\usepackage{epsfig}
\usepackage{graphicx}
\usepackage[dvipsnames]{xcolor}
\definecolor{linkcolor}{rgb}{0.0,0.3,0.5}
\usepackage{hyperref}
\usepackage{xspace}

\usepackage{cancel}
\usepackage{balance}
\usepackage[capitalise]{cleveref}
\usepackage{xspace}
\usepackage{enumerate}
\usepackage{ulem}
\usepackage{mdframed}
\usepackage{booktabs}
\usepackage{makecell}
\usepackage{enumitem,tabularx,multirow}
\usepackage{aas_macros}
\usepackage{colortbl}
\usepackage{orcidlink}

\AtBeginDocument{%
	\heavyrulewidth=.08em
	\lightrulewidth=.05em
	\cmidrulewidth=.03em
	\belowrulesep=.65ex
	\belowbottomsep=0pt
	\aboverulesep=.4ex
	\abovetopsep=0pt
	\cmidrulesep=\doublerulesep
	\cmidrulekern=.5em
	\defaultaddspace=.5em
}

\graphicspath{{Figures/}}

\usepackage{color}

\newcommand{\ba}{\begin{eqnarray}} 
	\newcommand{\ea}{\end{eqnarray}}  

\def\lsim{\mathrel {\vcenter {\baselineskip 0pt \kern 0pt \hbox{$<$} \kern 0pt \hbox{$\sim$} }}}

\def\gsim{\mathrel {\vcenter {\baselineskip 0pt \kern 0pt \hbox{$>$} \kern 0pt \hbox{$\sim$} }}}

\definecolor{mypurple}{RGB}{143, 116, 210}

\def\-{\,-\,}
\def\={\,=\,}
\def\+{\,+\,}

\definecolor{burgundy}{rgb}{0.5, 0.0, 0.13}
\definecolor{coolblack}{rgb}{0.0, 0.18, 0.39}
\definecolor{darkblue}{rgb}{0.0, 0.0, 0.55}
\definecolor{darkgreen}{rgb}{0.0, 0.2, 0.13}

\begin{document}
	\begin{frontmatter}
		\title{ Calculation of the transport coefficients in neutron star   }

		\author[1,2]{Utsab Gangopadhyaya,\orcidlink{0000-0002-9511-5817}}
		\ead{utsabgang@gmail.com}
		\author[1,2]{Suman Pal,\orcidlink{0009-0000-5944-4261}}
		\ead{sumanvecc@gmail.com}
		\author[1,2]{Gargi Chaudhuri,\orcidlink{0000-0002-8913-0658}}
		\ead{gargi@vecc.gov.in}
		\address[1]{Physics Group, Variable Energy Cyclotron Centre, 1/AF Bidhan Nagar, Kolkata 700064, India}
		
		\address[2]{Homi Bhabha National Institute, Training School Complex, Anushakti Nagar, Mumbai 400085, India}

		\begin{abstract}
			In this work, we have calculated the transport coefficients: shear viscosity and thermal conductivity inside the neutron star core. Our calculation is based on the relativistic kinetic theory approach using a modified BUU equation for quasi-particles whose mass and the chemical-potential and thus in turn the Fermi surface varies with the baryonic density $\rho_{B}$ and the temperature of the medium, and we have used the relaxation time approximation. For the description of the hadronic matter inside the neutron star, we consider the relativistic mean field model with three different kinds of parameterizations. We have found that the shear viscosity is predominantly influenced by neutrons, while thermal conductivity is primarily dominated by electrons.

		\end{abstract}
		
	\end{frontmatter}

	\section{Introduction} \label{sect:intro}
	
	The study of the physics and astrophysics of neutron stars \cite{Antoniadis:2013pzd, Fonseca:2021wxt, Riley:2021pdl, Miller:2021qha, Riley:2019yda, LIGOScientific:2017vwq, Radice:2017lry} provides a unique opportunity to analyze matter under extreme conditions.
	The recent detection of gravitational waves from binary neutron star mergers by the LIGO-Virgo collaboration offers a new context for examining the properties of super-dense, strongly interacting matter. In the post-merger remnant, there are significant density
	oscillations \cite{Alford:2017rxf,Alford:2019qtm} that can generate
	observable gravitational waves.  The oscillations and dissipative processes in post-merger neutron stars (NSs) are modeled using relativistic hydrodynamics, with the rate of dissipation being governed by the transport coefficients. 
	
	The exact structure and equation of state (EOS) of neutron stars (NSs) are still unknown and hence constitute an important subject of research in recent years \cite{Baym:2017whm, Lovato:2022vgq,Annala:2019puf}.The dynamics of non-equilibrium processes inside neutron stars are influenced by their internal structure, which in turn is reflected in key astrophysical observations. Thus, transport coefficients, which parametrize these non-equilibrium processes, help us understand the structure of NSs. The transport coefficients depend on the interactions of the constituent particles, the EOS, and other thermodynamic variables of the medium. Therefore, a correct choice of the internal structure ensures that the results of numerical simulations align with astrophysical observations. Various transport coefficients influence different astrophysical observables in distinct ways. For example, shear viscosity plays an important role in neutron star physics, it may damp the r-mode instability below a temperature of $10^8$K. Knowledge of shear viscosity is essential for understanding the pulsar glitches. The shear viscosity of neutron star matter was studied  by Flowers and Itoh \cite{Itoh}  and their results were used later by Cutler and Lindblom \cite{Cutler} to estimate the effects of dissipation on gravitational-wave instability. 
	Thermal conductivity is crucial for understanding the thermal structure and evolution of stars, particularly in modeling their cooling process. It is especially significant when examining the cooling of young neutron stars, where internal thermal relaxation is not yet fully attained.

	In this work, we intend to study shear viscosity and thermal conductivity.
	Neutron Stars as per current understanding, are composed of a dense core of asymmetric nuclear matter surrounded by a crust. The outer core mainly consists of neutrons, a small fraction of protons, and charge-neutralizing particles like electrons and muons.  The inner core, in addition to neutrons and protons, may also contain hyperons, kaon or pion condensates, and quarks. In this work, we will confine ourselves to a simplified model of the neutron star core consisting of neutrons, protons, electrons, and muons. The components are assumed to form a beta-stable nuclear matter, i.e., they are in beta equilibrium.
	For the description of the hadronic matter, we consider the Relativistic mean-field theory (RMF)\cite{glendenning2012compact}.
	
	We have employed a relativistic kinetic theory approach \cite{Groot,Prakash:1993bt,Albright:2015fpa,Gangopadhyaya:2016jrj,Gangopadhyaya:2017czt,Kalikotay:2019fle,Dash:2020vxk} to calculate the shear viscosity and thermal conductivity of neutron star matter. To solve the relativistic Boltzmann-Uehling-Uhlenbeck (BUU) transport equation, which governs the evolution of neutron star matter, we have used the relaxation time approximation. To calculate the relaxation time, we require the cross-sections for the interactions among the constituent particles. For nucleons, we have used a parametrized cross-section, while for leptons, we have used a cross-section calculated using a field-theoretic approach.  There are many recent works that have employed similar approach in order to derive the transport coefficients. They have taken into consideration quasi-particles as well as super fluidity in their calculations. They have used these values and have studied r-mode stability and evolution of the neutron star \cite{Cutler,Lindblom:2000gu,Laskos-Patkos:2023cts}. In some of these transport coefficient calculation they have either started from Landau's theory of Fermi liquid \cite{Itoh,flowers1979transport} or have started from BUU equation directly \cite{Baiko:2001cj,Shternin:2008es}. However, recent developments in the field of relativistic kinetic theory \cite{Albright:2015fpa} have shown that the introduction of an effective mass and effective chemical potential greatly alters the structure of the Boltzmann equation and the irreversible part of the energy momentum tensor. In this paper, we have taken such changes into consideration.

	This paper is organized as follows. In Sec.~\ref{sec:forma}, we provide a detailed description of the formalism used to calculate the transport coefficients of neutron star matter using relativistic kinetic theory. In Sec.~\ref{sec:results} we show the numerical results. Finally, we summarise our findings in Sec.~\ref{sec:conclusion}.

	\section{Formalism}\label{sec:forma}

	\subsection{Relativistic Mean Field Thory}

	We consider the RMF model for describing hadronic matter. In this model, the Lagrangian is given by: 
	
	\ba
	\mathcal{L}&=&\bar{\psi}(i\gamma_{\mu}\partial^{\mu}-m_N)\psi+\frac{1}{2}(\partial_{\mu}\sigma\partial^{\mu}\sigma-m_{\sigma}^2\sigma^2)-\frac{1}{4}\omega_{\mu\nu}\omega^{\mu\nu} \nonumber\\
	&+&\frac{1}{2}m_{\omega}^2\omega_{\mu}\omega^{\mu}-\frac{1}{4}\rho_{\mu\nu}\rho^{\mu\nu}+\frac{1}{2}m_{\rho}^2\vec{\rho_{\mu}}\vec{\rho^{\mu}}
	\nonumber\\
	&+& (g_{\sigma}\bar{\psi}\sigma\psi -g_{\omega}\bar{\psi}\gamma_{\mu}\omega^{\mu}\psi -\frac{1}{2} g_{\rho}\bar{\psi}\gamma_{\mu}\vec{\tau}.\vec{\rho^{\mu}})-\frac{1}{3}bm(g_{\sigma}\sigma)^3       \nonumber\\
	&-&\frac{c}{4} (g_{\sigma}\sigma)^4  +\Lambda_{\omega}g_{\omega}^2(\omega_{\mu}\omega^{\mu})(g_{\rho}^2\vec\rho_{\mu}\vec{\rho^{\mu}})\nonumber\\
	&+&\sum_{l= e,\mu}\bar{\psi_l}(i\gamma_{\mu}\partial^{\mu}-m_l)\psi_l  ~.
	\ea
	
	In this context, $\psi$ represents the nucleonic field, while $\psi_l$ denotes the leptonic field. The fields $\sigma$, $\omega_{\mu}$, and $\rho_{\mu}$ correspond to the scalar, vector, and isovector meson fields, respectively. The symbols $m_N$, $m_l$, $m_{\sigma}$, $m_{\omega}$, and $m_{\rho}$ refer to the masses of the nucleon, lepton, $\sigma$ meson, $\omega$ meson, and $\rho$ meson, respectively. The coupling constant $g_{\sigma}$ and $g_{\omega}$, along with $b$ and $c$ are determined by fixing the binding energy per nucleon ($B/A$), the nuclear incompressibility coefficient ($K_{sat}$), and the nucleonic effective mass ($m^*$) at the saturation densities ($\rho_0$). On the other hand, the coupling constants $g_{\rho}$ and $\Lambda_{\omega}$ are determined as function of symmetry energy ($E_{sym}$) and slope of the symmetry energy ($L_{sym}$) at the saturation density. 
	
	To simplify the solution of the field equations, we consider the mean-field approximation. We assume a static, uniform baryonic system in its ground state, where the meson fields are replaced by their expectation values. Since the meson fields are independent of spatial coordinates, the temporal and spatial derivatives of the fields vanish. However, the baryonic field operators remain unchanged. The Euler-Lagrange equation after applying the mean-field approximation, are given by:
	
	\ba
	m_{\sigma}^2{\sigma_0}  &=&-b m g_{\sigma}(g_{\sigma}\sigma_0)^2-cg_{\sigma}(g_{\sigma}\sigma_0)^3+\sum_{a=p,n} g_{\sigma }\gamma_{a} \int d\Gamma_a^* \frac{m_a^*}{E_a^*} f_a ~,  \nonumber\\ 
	m_{\omega}^2 \omega_0 &=& \sum_{a=p,n} g_{\omega }\gamma_{a} \int  d\Gamma_a^*  f_a  ~,  \nonumber \\ 
	m_{\rho}^2 \rho_{30} &=& \sum_{a=p,n} g_{\rho }\gamma_{a} \int d\Gamma_a^*  f_a I_{3a} ~. 
	\ea
	
	Here, $\sigma_0$, $\omega_0$, and $\rho_{30}$ are the space-time independent mean field values; $d\Gamma_a^* =\frac{ d^{3}p_{a}^*}{(2\pi)^3}$; $\gamma_a$ is the degeneracy factor (equal to 2 for both nucleons and leptons), $I_{3a}$ represents the isospin of the nucleons($I_{3a}=+1/2$ for proton and $I_{3a}=-1/2$ for neutron).
	In a uniform medium at thermal equilibrium, the meson fields develop constant, nonzero mean values, represented by $\bar{\sigma}$, $\bar{\omega}^{\mu}$, and $\bar{\rho}^{\mu}$, which do not vary with space or time. In the rest frame of the medium, the spatial components of the vector fields vanish due to rotational symmetry. However, in a more general frame of reference, these spatial components do not necessarily vanish.
	Here $m_a^*=m_a-g_\sigma\bar{\sigma}$ is the effective mass (in the case of leptons, $m_l^{*}=m_l$, where $m_l$ is the leptonic mass.). The four-momentum is given by $p_{a}^{\mu}=p_{a}^{* \mu}+g_{\omega} \bar{\omega}^{\mu}+g_{\rho}I_{3a}\bar{\rho^{\mu}}$, and the effective chemical potential is $\mu_a=\mu_a^*+g_{\omega}\bar{\omega^0}+g_{\rho}\bar{\rho_{3}^0}I_{3a}=\mu_B-q_{a}\mu_e$. The effective energy is $E_a^*=\sqrt{(p_{a}^{*})^2+(m_{a}^*)^{2}}$, and the total energy $E_{a}=E_{a}^{*}+g_{\omega}\bar{\omega^0}+g_{\rho}\bar{\rho_{3}^0}I_{3a}$, where $I_{3a}$ represents the iso-spin.
	In equilibrium the distribution function for a particle of type $a$ is given by:
	\begin{equation}
		f_{a}=\frac{1}{1+\exp(\frac{E_a^*-\mu_a*}{T})}. 
		\label{eq:Fermi_Dirac}
	\end{equation}
	The energy-momentum tensor describes the internal properties of a matter 
	distribution and it arises from the space-time symmetry. Using infinitesimal space-time transformations, we can derive the energy-momentum tensor as 
	\begin{equation}
		T^{\mu\nu}=-g^{\mu\nu}\mathcal{L}+\sum\frac{\partial\mathcal{L}}{\partial(\partial_{\mu}\phi)}\partial^{\nu}\phi ,
	\end{equation} 
	where $\phi$ is an arbitrary field (fermionic, meson, etc.). As in the RMF model, we consider the fields $\psi,\bar{\psi},\sigma,\omega, \rho $. The Lagrangian after applying the mean-field approximation reads: 
	\ba
	\mathcal{L}_{RMF} &=& -\frac{1}{2}m_{\sigma}^2{\bar{\sigma}}^2+\frac{1}{2}m_{\omega}^2(\bar{\omega}^0)^2+\frac{1}{2}m_{\rho}^2(\bar{\rho}_{3}^0)^2     \nonumber\\
	&+& \Lambda_{\omega}(g_{\omega}\bar{\omega}^0)^2(g_{\rho}\bar{\rho}_{3}^0)^2-\frac{1}{3}bm\big(g_{\sigma}\bar{\sigma}\big)^3-\frac{c}{4}\big(g_{\sigma}\bar{\sigma}\big)^4. ~~~
	\ea
	The energy-momentum tensor is expressed as
	\begin{equation}
		T^{\mu\nu}=g^{\mu\nu}U+\sum_a \gamma_{a} \int d\Gamma_a^* \frac{p_a^{\mu} p_a^{*\nu}}{E_a^*} f_a  .   \label{eq:tot_EMtensor}
	\end{equation}
	Where $U=-\mathcal{L}_{RMF}$
	The baryon current is given by
	\ba \label{eq:current} 
	J_B^{\mu}=\sum_a b_{a}\gamma_{a}\int d\Gamma_a^* \frac{ p_a^{*\mu}}{E_a^*} f_a  .
	\ea


	\subsection{Quasi particle transport theory}

	The general Boltzmann equation \cite{Albright:2015fpa} is given by the following relation,
	\ba 
	\frac{df}{dt}=\mathbb{C}_{a} .
	\label{eq:BE_1}
	\ea 
	Which for a quasi-particle formalism takes the form,
	
	\ba
	\frac{df_a}{dt}({\bf x},{\bf p}^*,t) = \frac{\partial f_a}{\partial t} + \frac{\partial f_a}{\partial x^i } \frac{d x^i}{dt} + \frac{\partial f_a}{\partial p^{*i}} \frac{d p^{*i}}{dt} &=&
	\mathbb{C}_a  ,
	\label{eq:BE_2}   \\
	- \left[ \frac{m_a^*}{E_a^*} \, \frac{\partial m_a^*}{\partial x^i} 
	+ g_{\omega a}  \frac{p_{\mu}^*}{E_a^*} \,  \bar{\omega}^{\mu i}+g_{\rho a} I_{3a} \frac{p_{\mu}^*}{E_a^*} \,  \bar{\rho}^{\mu i} \right] \frac{\partial f_a}{\partial p^{*i}} 
	+\frac{p^{*\mu}}{E_a^*} \partial_{\mu} f_a &=& \mathbb{C}_a .~~
	\label{eq:BE_3}
	\ea
	
	Here we have used the expression of velocity in terms of the momentum and the energy 
	\ba
	\frac{d x^i}{dt} = \frac{\partial E_a}{\partial p_a^i} = \frac{p^{*i}}{E_a^*} \,. \label{velocity}
	\ea
	as well as the relativistic version of Newton's Second Law
	\ba
	\frac{d p_a^i}{d t} &=& - \left( \frac{\partial E_a}{\partial x^i} \right)_p   \nonumber \\
	&=&-\Big( \frac{m_a^*}{E_a^*} \, \frac{\partial m_a^*}{\partial x^i} - g_{\omega a}  \frac{\partial \bar{\omega}^j}{\partial x^i}
	\frac{p^{*j}}{E_a^*} + g_{\omega a} \frac{\partial \bar{\omega}^0}{\partial x^i}  - I_{3a}g_{\rho a}  \frac{\partial \bar{\rho}^j}{\partial x^i}
	\frac{p^{*j}}{E_a^*} + I_{3a} g_{\rho a} \frac{\partial \bar{\rho_{3}}^0}{\partial x^i}
	\Big) \,.
	\ea

	The general form of the energy-momentum tensor ($T^{\mu\nu}$) and baryon current ($J_B^{\mu}$) in terms of the energy density $\varepsilon$, pressure $P$, baryon density $\rho_{B}$ in the local rest frame, along with the four velocity  $u^{\mu}$, is given by:
	\ba
	T^{\mu\nu}&=&-Pg^{\mu\nu}+(\varepsilon+P)u^{\mu}u^{\nu}+\Delta T^{\mu\nu}, \label{eq:Tmunu}\\ 
	J_B^{\mu}&=&\rho_B u^{\mu}+\Delta J_B^{\mu}. \label{eq:jmu}
	\ea
	The common thermodynamic relations between the parameters defined in the local rest frame holds true,
	\ba
	s &=& \partial P/\partial T    ,  \label{eqn:entropy}\\
	\rho_{B} &=& \partial P/\partial \mu_{B}, \\
	\varepsilon &=& -P + Ts + \mu_B \rho_B,
	\ea
	where $s$, $T$, $\mu_{B}$, and $\rho_{B}$ denote the entropy density, temperature, and baryonic chemical potential, and baryonic density, respectively. Here in this paper, we will follow the Landau definition of four-velocity $u^{\mu}$ which is parallel to the energy transport. Accordingly: 
	\ba
	\Delta T^{\mu\nu} &=& \eta \left( \nabla^{\mu} u^{\nu} + \nabla^{\nu} u^{\mu} + \frac{2}{3} \Delta^{\mu\nu} \partial \cdot u \right)   
	- \zeta \Delta^{\mu\nu} \partial \cdot u  ~,  
	\label{eq:DTuv}  \\
	\Delta J_B^{\mu} &=& \lambda \left( \frac{\rho_B T}{w}\right)^2 \nabla^{\mu} \left( \frac{\mu_B}{T} \right) \,.
	\label{eq:DJB}
	\ea
	where $w$ is the enthalpy density, and
	\ba
	\nabla^{\mu}&=&\partial^{\mu}-u^{\mu}D ;~~~D=u^{\mu}\partial_{\mu} ;\\
	\Delta^{\mu\nu}&=&u^{\mu}u^{\nu}-g^{\mu\nu}.
	\ea
	
	Using the BUU transport equation, the conservation equations for energy, momentum, and baryon current can be derived,
	\ba 
	\partial_{\mu} T^{\mu\nu} &=& 0,   \\
	\partial_{\mu} J_B^{\mu} &=& 0 .
	\ea 
	To derive these, we use the constraint:
	\ba
	\sum_a \int d\Gamma_a^* \chi_a \mathbb{C}_a = 0 \, ,
	\ea
	where $ \chi_a$ represents any quantity of the quasi-particle $a$ that is conserved during collision or decay (i.e. energy, momentum, or baryon number).
	
	For a first-order departure from equilibrium, the distribution function can be written as follows:
	\begin{equation}
		f_{a}=f_{a}^{0}+\delta \tilde{f}_{a}=f_{a}^{0}+f_{a}^{0}(1-f_{a}^{0})\phi_{a}.
	\end{equation} 
	Here $f_{a}^{0}$ represents the equilibrium distribution function, while $\phi_{a}$ parametrizes the deviation from equilibrium, serving as the sole contributor to the dissipative part ($\Delta T^{\mu\nu}$ and $J_B^{\mu}$).
	The form of $\phi_{a}$ follows from the dissipation part as: 
	\ba
	\phi_{a}=A_{a}\partial_{\mu}u^{\mu} + B_{a} p_{a}^{\mu} \nabla_{\mu}\left(\frac{\mu_B}{T}\right)       
	- C_{a} p_{a}^{\mu} p_{a}^{\nu} \left( \nabla_{\mu} u_{\nu} + \nabla_{\nu} u_{\mu} - \frac{2}{3} \Delta_{\mu\nu} \partial_{\rho} u^{\rho} \right),   \label{eq:phi_a}
	\ea
	where the coefficients $A_{a}$,$B_{a}$,~and ~$C_{a}$~depends only on the momentum $p$ and the species of the particle, where as $u^{\mu}$ depends only on the space-time co-ordinates x.  
	
	In systems where the effective mass is not taken into consideration the distribution function of the system slightly away from equilibrium is expanded around the equilibrium energy $E_{a}^{0}$, 
	
	\ba
	f_a(E_a, T, \mu_B) &=& f_a^{0}(E_a^0, T^0, \mu_B^0) + \delta f_a . 
	\ea
	
	where $\delta f_a=f_{a}^{0}(1-f_{a}^{0})\phi_{a}$, and $\phi_{a}$ is the same as defined in Eqn.(\ref{eq:phi_a}). However, for a quasiparticle distribution function, since the effective mass and other mean-field terms varies with the thermodynamic parameters, $\delta f_{a}$ aquires an extra term involving $\delta E_{a}$.
	\ba
	\delta E_{a}=\frac{m_{a}^{*}}{E_{a}^{*}} \delta m_{a}^{*}+g_{\omega}\delta\bar{\omega^0}+g_{\rho}I_{3a}~\delta\bar{\rho_{3}^0}.
	\ea
	If we expand the quasiparticle distribution function around the non equilibrium energy $E_{a}$, that is conserved in a collision we get,
	\ba
	f_a(E_a, T, \mu_B) &=& f_a^{0}(E_a, T^0, \mu_B^0) + \delta \tilde{f}_a \,.
	\ea
	The terms $\delta f_a$ and $\delta \tilde{f}_a$ are related to each other by,
	\ba
	\delta f_a = \delta \tilde{f}_a + \left( \frac{\partial f_a^{0}}{\partial E_a} \right)_{T^0 ,\, \mu_B^0} \delta E_a
	= \delta \tilde{f}_a -  f_a^{0}\left(1-f_a^{0}\right) \frac{\delta E_a}{T}.
	\label{ftildef}
	\ea
	Now if we try to calculate the energy momentum tensor resulting from departure from equilibrium we will find that the expression,
	
	\ba
	\delta T^{ij}=\sum_{a} \gamma_{a}\int d\Gamma_{a}^{*} \frac{p_{a}^{*i}p_{a}^{*j}}{E_{a}^{*}}\left(\delta f_{a}-f_{a}^{0}\frac{\delta E_{a}^{*}}{E_{a}^{*}}\right)+g^{ij}\delta U    \label{eq:Tij1}
	\ea 
	
	simplifies to,
	\ba 
	\delta T^{ij} &=& \sum_a \gamma_{a}\int d\Gamma_a^{*}\frac{p_a^ip_a^j}{E_a^*}\delta \tilde{f}^{a}.   \label{eq:Tij}
	\ea
	The above equation has been derived from Eqn.~(\ref{eq:Tij1}) by taking into consideration the fact that the mean field contains no entropy, and $\delta T$, $\delta \mu_{B}$ are not independent, thus:
	
	\ba
	\delta U&=&-\sum_{a}\gamma_{a}\int d\Gamma_{a}^{*} \delta E_{a}f_{a}^{0}, \\
	f_a^{0}\left(1-f_a^{0}\right) \frac{\delta E_a}{T}&=& \left[ \frac{ T( \partial E_a/\partial T)_{\sigma}}{E_a - \mu_a + T \left( \partial \mu_a / \partial T \right)_{\sigma} } \right] \delta \tilde{f}_a .
	\ea
	
	Similarly, other components of the energy-momentum tensor and the baryonic current are given by:
	
	\ba
	\delta T^{00} &=& \sum_{a} \gamma_{a}\int d\Gamma_{a}^{*} E_{a}\left\{1-\frac{T(\frac{\partial E_a}{\partial T})_{\sigma}}{E_a-\mu_a+T(\frac{\partial \mu_a}{\partial T})_{\sigma}}\right\}\delta\tilde{f}^{a}  , ~~ ~    \label{eq:T00}    \\
	\delta T^{0j} &=& \sum_{a} \gamma_{a}\int d \Gamma_{a}^{*} \frac{p_a^{*j}}{E^{*}_a}E_{a} \delta f_a=\sum_{a} \gamma_{a}\int d \Gamma_{a}^{*} \frac{p_a^{*j}}{E^{*}_a}E_{a} \delta \tilde{f_a} , \label{eq:T_0j}~~~  \\
	\delta J_{B}^{i}&=& \sum_{a} b_{a} \gamma_{a} \int d \Gamma_{a}^{*} \frac{p_a^{*j}}{E^{*}_a} \delta \tilde{f}_a .  \label{eq:JiB}
	\ea

	The detailed derivations leading to Eqs.~(\ref{eq:Tij})-(\ref{eq:JiB}) can be found in\cite{Albright:2015fpa}. The extreme right hand side of the Eqn.~(\ref{eq:T_0j}) can be obtained from the middle part by noting, that the term that differentiates $\delta f_{a}$ from $\delta \tilde{f}_{a}$ as expressed in Eqn.~(\ref{ftildef}) is spherically symmetric in momentum space, and thus integrates to zero. 
	
	Comparing Eqn.~(\ref{eq:Tij}) and Eqn.~(\ref{eq:JiB}) with Eqn.~(\ref{eq:DTuv}) and Eqn.~(\ref{eq:DJB}), we obtain the required expression for the transport coefficients in terms of coefficients of $\phi_{a}$ \cite{Albright:2015fpa}:

	\begin{equation}
		\begin{aligned}
			\eta=&\frac{2}{15}\sum_a \gamma_{a} \int d\Gamma_a^{*} \frac{ |{\bf p}_a^*|^4}{E^{*}_a}f^{0}_{a}\left(1-f^{0}_{a}\right) C_{a},\\
			\lambda=&\frac{1}{3}\left(\frac{w}{\rho_{b}T}\right)^{2}\sum_a b_{a} \gamma_{a}\int d\Gamma^{*}_a \frac{ |{\bf p}_a^*|^2}{E^{*}_a}f^{0}_{a}\left(1-f^{0}_{a}\right)
			B_{a}.  \\
		\end{aligned}
	\end{equation}

	\subsection*{Relaxation time approximation}
	
	The term on the right hand side of Eqn.~(\ref{eq:BE_1}) is known as the collision integral, and its explicit form for 2-to-2, 2-to-1 and 1-to-2 processes is given by~\cite{Albright:2015fpa,Groot}: 
	\ba
	\mathbb{C}_a 
	&=& \sum_{bcd} \frac{\gamma_{b}}{1+\delta_{ab}}
	\int d\Gamma_b^* \, d\Gamma_c^* \, d\Gamma_d^* \,
	W(a,b|c,d)   
	\{f_c f_d\left(1-f_a\right)\left(1-f_b\right) - f_a f_b\left(1-f_c\right)\left(1-f_d\right)\}   \nonumber \\
	&
	+& \sum_{cd} \int d\Gamma_c^* \, d\Gamma_d^* \, W(a|c,d)   
	\{f_c f_d\left(1-f_a\right) - f_a\left(1-f_c\right)\left(1-f_d\right)\}
	\nonumber \\
	&
	+& \sum_{bc} \int d\Gamma_b^* \, d\Gamma_c^* \, W(c|a,b)    
	\{f_c\left(1-f_a\right)\left(1-f_b\right) - f_a f_b\left(1-f_c\right)\} \,.
	\label{eq:collisionintegral:collision_full1}
	\ea
	where,
	\ba 
	W(a,b|c,d) &=& \frac{(2\pi)^4 \delta^4(p_a + p_b - p_c - p_d ) }
	{2E_a^* 2E_b^* 2E_c^* 2E_d^*} 
	| \overline{\mathcal{M}(a,b|c,d)} |^2  ,  ~~~~~
	\label{eq:collisionintegral:W2to2}   \\
	W(a|c,d) &=& \frac{(2\pi)^4 \delta^4(p_a - p_c - p_d ) }
	{2E_a^* 2E_c^* 2E_d^*} 
	| \overline{ \mathcal{M}(a|c,d) } |^2  ;
	\label{eq:collisionintegral:W1to2}
	\ea 
	and~\cite{Palash,Prakash:1993bt},
	\ba
	\frac{d\sigma_{a,b\rightarrow c,d}}{d\Omega}  &=& \frac{1}{64\pi^{2}s}\frac{p_{f}}{p_{i}} | \overline{\mathcal{M}(a,b|c,d)} |^2,     \\
	\Gamma_{a\rightarrow~c,d} &=& \frac{p_{f}}{8\pi \left(m_{a}^{*}\right)^{2}}  | \overline{ \mathcal{M}(a|c,d) } |^2. 
	\ea
	
	In the centre of mass frame $p_{f}=p_{i}$ and, $s$ is the square of the total centre of mass energy (Mandelstam variable).
	Here $|\overline{\mathcal{M}}|^{2}$ has been summed over all final, and averaged over initial spin states. Additionally,  $E_{a}^{*}$ has been used instead of $E_{a}$ in the denominators of the transition rates $W$ to make the phase-space integration Lorentz covariant.
	
	In the relaxation time approximation method, the collision integral is replaced with $-\frac{\delta f_{a}}{\tau_{a}}$ in the Boltzmann equation~\cite{Landau,Prakash:1993bt},
	\ba
	\frac{d f_{a}^{0}}{dt}=\mathbb{C}_{a}=-\frac{f_{a}^{0}\left(1-f_{a}^{0}\right)\phi_a}{\tau_a}.   \label{eq:RT_Colli_Integral}
	\ea 
	
	To derive the relaxation time for a particle of species `$a$' from the scattering amplitudes of individual processes, we replace the distribution function of particle `$a$', $f_{a} \rightarrow f_{a}^{0} + \delta f_{a}$, and assume the distribution functions of other particles to be at their respective equilibrium states~\cite{Prakash:1993bt,Gangopadhyaya:2016jrj,Gangopadhyaya:2017czt,Kalikotay:2019fle,Dash:2020vxk}. Thus,
	\ba
	\frac{1}{\tau_a(E_a^*)} 
	&=&  \sum_{bcd} \frac{\gamma_{b}}{1+\delta_{ab}}
	\int d\Gamma_b^* \, d\Gamma_c^* \, d\Gamma_d^* \,   
	W(a,b|c,d)  
	\frac{f_c^{0}f_d^{0}}{f_a^{0}}\left(1-f_b^{0}\right)   \nonumber \\
	&+&  \sum_{cd} \int d\Gamma_c^* \, d\Gamma_d^* \, W(a|c,d) \frac{f_c^{0}f_d^{0}}{f_a^{0}} 
	\nonumber \\
	&+& \sum_{bc} \int d\Gamma_b^* \, d\Gamma_c^* \,
	f_b^{\rm eq} W(c|a,b) \frac{f_c^{0}\left(1-f_b^{0}\right)}{f_a^{0}}  . ~~~
	\label{eq:RTA:relaxationtime}
	\ea
	To arrive at the above equation, we used the relations 
	$f_c^{0} f_d^{0}\left(1-f_a^{0}\right)= f_a^{0} \left(1-f_c^{0}\right)\left(1-f_d^{0}\right)$
	and  
	$f_c^{0} f_d^{0}\left(1-f_a^{0}\right)\left(1-f_b^{0}\right) = f_a^{0} f_b^{0}\left(1-f_c^{0}\right)\left(1-f_d^{0}\right)$ 
	for processes $a+b \rightarrow c+d$ and $a \rightarrow c+d$ respectively. 
	
	The system under consideration consists only of fermions, and hence the ratio $\frac{\mu_{a}^{}}{T}$ for any particle is extremely small. As a result, all the states below the Fermi energy are almost completely filled, while those above are nearly empty. Particles with energy below the Fermi energy cannot interact, as the final states of any such possible interaction are fully occupied, and there are practically no particles with energy greater than the Fermi energy available to interact. Similarly, there is no decay or inverse decay for particles with energy less than the Fermi energy, as the energy of the produced particles would fall below their respective Fermi energies; this is ensured by conditions like $\mu_{N}^{*}\approxeq \mu_{P}^{*}+\mu_{e}^{*}$. Therefore, the only particles of interest are those within a narrow band around the Fermi energy of the respective particles, the width of which is determined by $T$. Thus, the only relaxation time of interest is that at the Fermi energy of the particle (i.e. $E^{*}_{a}=\mu^{*}_{a}$).

	\ba
	\frac{1}{\tau_a(\mu_a^*)} 
	&=&  \sum_{bcd} \frac{\gamma_{b} T}{1+\delta_{ab}}\frac{\sqrt{\left(\mu_{b}^{*}\right)^{2}-\left(m_{b}^{*}\right)^{2}}}{\left(2\pi\right)^{2}\mu_{k}^{*}}
	\int d\left(\cos{\Theta}\right) \frac{\sqrt{\lambda\left(s,\left(m_{a}^{*}\right)^{2},\left(m_{b}^{*}\right)^{2}\right)}}{2} \,\sigma_{a,b\rightarrow c,d}\left(s\right)  \nonumber \\
	&+&  \sum_{cd} \frac{m_{a}^{*}}{2\mu_{a}^{*}}\Gamma_{a\rightarrow c,d}
	+  \sum_{bc} \frac{\sqrt{\left(p_{f}\right)^{2}+\left(m_{a}^{*}\right)^{2}}}{2\mu_{a}^{*}} \Gamma_{c\rightarrow a,b}   ~~.
	\label{eq:RTA:relaxationtime2}
	\ea
	where,
	\ba
	s &=&\left(m_{a}^{*}\right)^{2} + \left(m_{b}^{*}\right)^{2} + 2 \mu^{*}_{a}\mu^{*}_{b} 
	- 2 \sqrt{\left(\mu_{a}^{*}\right)^{2}-\left(m_{a}^{*}\right)^{2}} 
	\cdot \sqrt{\left(\mu_{a}^{*}\right)^{2}-\left(m_{a}^{*}\right)^{2}}\cos{\Theta} ,  \\
	p_{f}&=&\frac{\sqrt{\lambda\left(\left(m_{a}^{*}\right)^{2},\left(m_{c}^{*}\right)^{2},\left(m_{d}^{*}\right)^{2}\right)}}{2m_{a}^{*}}.
	\ea 
	The function $\lambda(x,y,z)=x^{2}+y^{2}+z^{2}-xy-yz-zx$ in the above equation is the K\"{a}ll\'{e}n function.  To obtain Eqn.~(\ref{eq:RTA:relaxationtime2}) we used the conditions \cite{Sen:2021tdu} $\frac{f^{0}\left(1-f^{0}\right)}{T}\approxeq \delta\left(E^{*}-\mu^{*}\right)$ and $f^{0} \approxeq \frac{1}{2}$.

	\subsection*{Final Formula for transport coefficient}

	The final  resulting expressions for the transport coefficients are (The details are extensively discussed in ~\cite{Albright:2015fpa}):     
	
	\ba
	\eta & =&\frac{1}{15T}\sum_a \gamma_{a} \int  d\Gamma_a^{*} \frac{|p_a^{*}|^4}{E_a^{*2}} \tau_a(E_a^*)  f_{a}^{0}\left(1-f_{a}^{0}\right) ,   \\
	\lambda &=&\frac{1}{3} \left(\frac{w}{\rho_{B}T}\right)^{2}  \sum_a \gamma_{a} \int d\Gamma_a^{*} \frac{|p_a^{*}|^2}{E_a^{*}} \tau_a(E_a^*) 
	\nonumber\\
	& & ~~~~~ \left(b_{a}-\frac{\rho_B \mu_a}{(\varepsilon+P)}\right)^2    f_{a}^{0}\left(1-f_{a}^{0}\right)   .  
	\ea
	
	Now for the system we are dealing with, $\frac{f^{0}\left(1-f^{0}\right)}{T}\approxeq \delta\left(E^{*}-\mu^{*}\right)$,  the above expressions take the following form, 
	\ba 
	\eta & =&\frac{1}{15}\sum_a \gamma_{a} \int  d\Gamma_a^{*} \frac{|p_a^{*}|^4}{E_a^{*2}} \tau_a(E_a^*) \delta\left(E_{a}^{*}-\mu_{a}^{*}\right) ~,  \label{eq:eta}   \\
	\lambda &=&\frac{1}{3T} \left(\frac{w}{\rho_{B}}\right)^{2}  \sum_a \gamma_{a} \int d\Gamma_a^{*} \frac{|p_a^{*}|^2}{E_a^{*2}} \tau_a(E_a^*) 
	\left(b_{a}-\frac{\rho_B \mu_a}{w}\right)^2    \delta\left(E_{a}^{*}-\mu_{a}^{*}\right)  ~.  \label{eq:lambda}
	\ea
	
	\begin{figure*}[htp]
		\centering
		\includegraphics[width=1.0\textwidth]{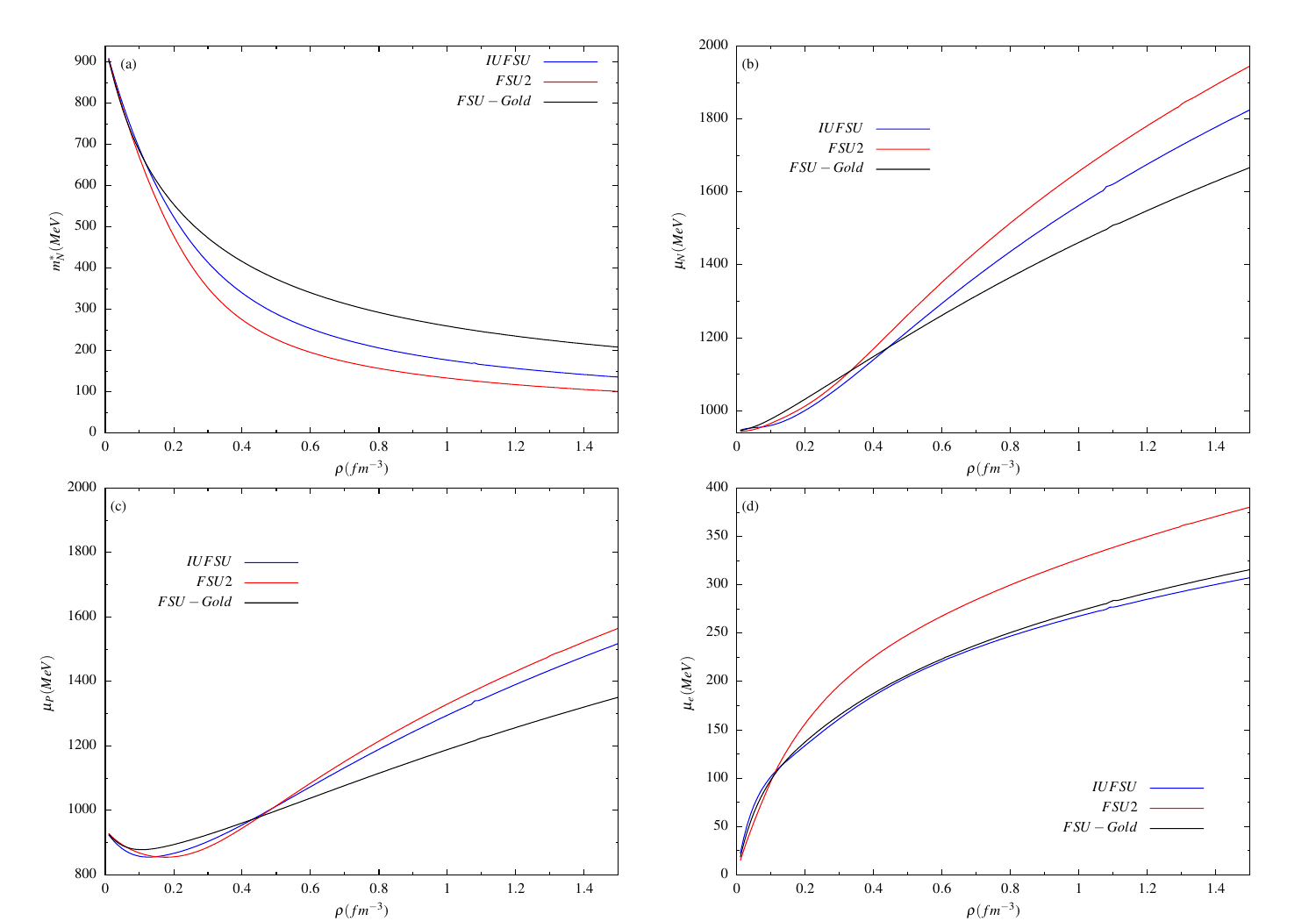}
		\caption{\it Vartion of the effective mass and chemical potential of the nucleon and lepton with density for different hadronic models}
		\label{fig:massmu}
	\end{figure*} 
	\begin{figure*}[htp]
		\centering
		\includegraphics[width=1.0\textwidth]{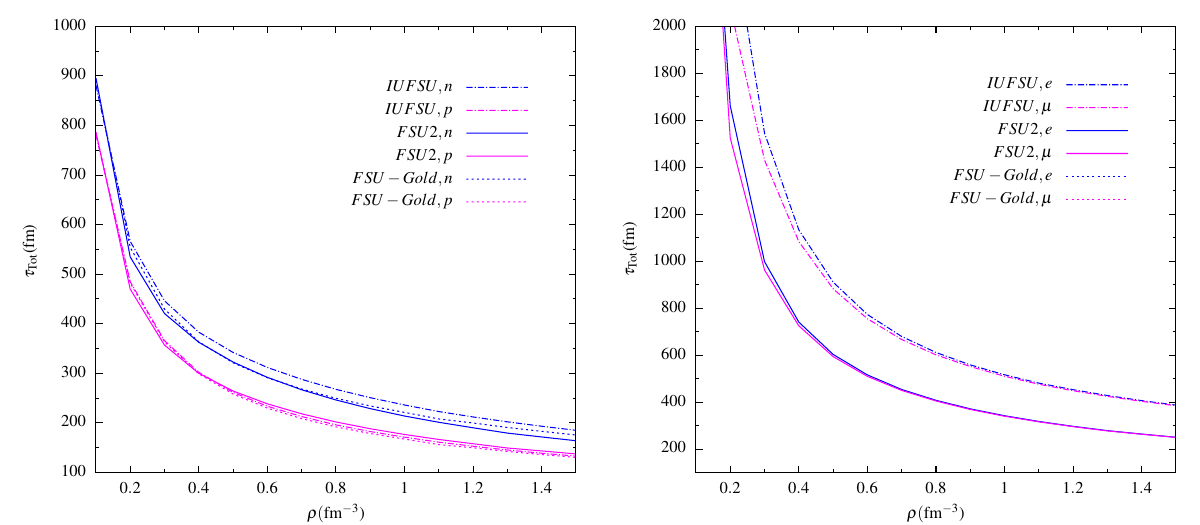}
		\caption{\it Total relaxation time for nucleon (a) and lepton (b) with density.}
		\label{fig:total_relax_time}
	\end{figure*}

	\section{Results}\label{sec:results} 
	
	In this study, we have determined the transport coefficients within neutron stars. The internal composition of the neutron star has been modeled using relativistic mean-field (RMF) approaches. We assume that the matter inside the neutron star is composed of neutrons, protons, electrons, and muons. Specifically, we employ three established variants of the RMF model: IUFSU \cite{Fattoyev:2010mx}, FSU2 \cite{Chen:2014sca}, and FSUGold \cite{Todd-Rutel:2005yzo}. The details of nuclear matter properties and the nucleon-meson coupling are given in Tables.~\ref{tab:1} and \ref{tab:2}.
	To derive the equation of state for a neutron star, we impose conditions such as charge neutrality, baryon number conservation, and beta equilibrium. To determine the transport coefficients, we require the nucleon effective mass, the chemical potentials of various particles, and the enthalpy, all of which were previously discussed in Section~\ref{sec:forma}.
	\begin{table}[!ht]
		\caption{The nuclear matter properties at saturation density for the three different hadronic models.}
		\setlength{\tabcolsep}{12.0pt}
		\begin{tabular}{ccccccc}
			\hline
			\hline
			Model&$\rho_{0}$ & $B/A$ & $K_{sat}$ & $m^*/m$ & $E_{sym}$& $L_{sym}$\\
			Name& $({\rm fm}^{-3})$ & (MeV) & (MeV) & ~& (MeV) & (MeV)\\ \hline
			IUFSU&0.155 & -16.40 & 231.2 & 0.61&31.30&47.2\\
			FSU Gold&0.148 & -16.28 & 230.0 & 0.61& 32.59&60.5\\
			FSU2&0.1505  & -16.28 & 238.0 & 0.593 & 37.62& 112.8\\
			\hline
			\hline
		\end{tabular}
		\label{tab:1}
	\end{table} 
	
	In Fig.~\ref{fig:massmu}, we show the variation of the effective mass and the different chemical potentials of the hadronic matter for three different RMF model parametrizations. A very low temperature (0.1 MeV) is used in this work. In Fig.~\ref{fig:massmu}(a), we show the variation of nucleon effective mass ($m_N^*$) with density. The effective equals $m_N$ at low density and it decreases at high density. In Fig.~\ref{fig:massmu}(b), we present the neutron chemical potential; in Fig.~\ref{fig:massmu}(c), the proton chemical potential, and in Fig.~\ref{fig:massmu}(d), the electron chemical potential. When the electron chemical potential exceeds muon mass, muons appear and from the $\beta$ equilibrium condition, muon chemical potential  equals  the electron chemical potential, hence we have not shown the muon chemical potential in this figure. \\
	The expressions for the transport coefficients of a multi-component Fermi gas, as discussed in the previous sections, are quite general. The only factors that differentiate one Fermi gas from another are the interactions between the constituent species and the ways in which these species decay. Information about the interactions is introduced through the cross-sections, while decay processes are characterized by the decay rates. For nucleon-nucleon collisions, we have used the parameterized form of the vacuum cross-section as mentioned in \cite{Prakash:1993bt,Bertsch:1988ik}. 
	\begin{table}[!ht]
		\caption{Model parameters used in the calculations. The masses of the nucleon($m_N$), omega ($m_{\omega}$), and rho ($m_{\rho}$) are fixed at 939 MeV, 782.5 MeV, and 763 MeV, respectively.}
		\setlength{\tabcolsep}{2 pt}
		\begin{tabular}{ccccccccc}
			\hline
			\hline
			Model&$m_{\sigma}(MeV)$ &$g_{\sigma}^2$ & $g_{\omega}^2$ & $g_{\rho}^2$ & $b$ & c& $\Lambda_{\omega}$ &$\zeta_{\omega}$\\
			\hline
			IUFSU&491.5&99.4266& 169.8349 & 184.6877 & 0.0018002 & 0.0000986& 0.046 &0.03\\
			FSU Gold&491.5& 112.1996 & 204.5469 & 138.4701 & 0.0007562 & 0.00793 & 0.03 &0.06\\
			FSU2&497.479&108.0943&183.7893&80.4656& 0.001598& 0.000177& 0.000823&0.0256\\
			\hline
		\end{tabular}
		\label{tab:2}
	\end{table}

	The expression of  $NN$ cross-section :
	\begin{equation} \label{eq:NN_crosssection}
		\sigma^{NN}(\sqrt{s}) = \begin{cases}
			5.5 ~\text{fm}^{2} & \text{if } \sqrt{s} < 1.8993 \\[10pt]
			\frac{3.5 \, \text{fm}^2}{1 + 100(\sqrt{s} - 1.8993 \, \text{GeV})} + 2.0 \, \text{fm}^2& \text{if } \sqrt{s} > 1.8993    
		\end{cases}  ~~.
	\end{equation} 
	We have neglected the interaction between the nucleons and leptons, as such interactions are not included in the RMF model; therefore, the respective cross-section is assumed to be zero. The $\beta$-decay rate is taken from \cite{Shapiro:1983du}, where the neutron decay rate is given as $\frac{1}{972}\,\text{s}^{-1}$. In this work, we have ignored Coulomb effects and radiative corrections, as mentioned in \cite{Shapiro:1983du}. For lepton-lepton scattering, we have calculated the cross-sections using QED. The cross-section diverges for scattering angles of $0$ and $\pi$, but if we account for the Debye mass \cite{Masood:2020sgp,debye}, these non-physical characteristics of the cross-section disappear. The amplitudes for leptonic scattering are given below:
	
	\ba 
	| \overline{ \mathcal{M}(e,e|e,e) } |^2  &=&\frac{2e^4}{32\pi s}\Big[\frac{1}{u_D^2}(s^2+t^2-8m_e^2(s+t)+24m_e^2)    \nonumber\\
	&+& \frac{1}{t_D^2}(s^2+u^2-8m_e^2(s+u)+24m_e^2)    +\frac{2}{u_Dt_D}(s^2-8m_e^2s+12m_e^4)\Big],~~~~  \label{eq:llcrosssection_1}\\ 
	| \overline{ \mathcal{M}(\mu,\mu|\mu,\mu) } |^2 &=&\frac{2e^4}{32\pi s}\Big[\frac{1}{u_D^2}(s^2+t^2-8m_{\mu}^2(s+t)+24m_{\mu}^2)    \nonumber\\
	&+&\frac{1}{t_D^2}(s^2+u^2-8m_{\mu}^2(s+u)+24m_{\mu}^2)   +\frac{2}{u_Dt_D}(s^2-8m_{\mu}^2s+12m_{\mu}^4)\Big],~~~\label{eq:llcrosssection_2}\\ 
	| \overline{ \mathcal{M}(e,\mu|e,\mu) } |^2  &=&\frac{2e^4}{t_D^2}(s^2+u^2-2(m_e^2+m_{\mu}^2)(s+u)+6(m_e^2+m_{\mu}^2)^2)~.\label{eq:llcrosssection_3}
	\ea
	In this context, \(s\), \(t\), and \(u\) represent the Mandelstam variables. As mentioned earlier, due to the presence of the Debye mass, the variables undergo modifications such that \(t \rightarrow t_D\) and \(u \rightarrow u_D\), where \(t_D = t - m_D\) and \(u_D = u - m_D\), with \(m_D\) being the Debye mass.

	\begin{figure*}[htp]
		\centering
		\includegraphics[width=1.0\textwidth]{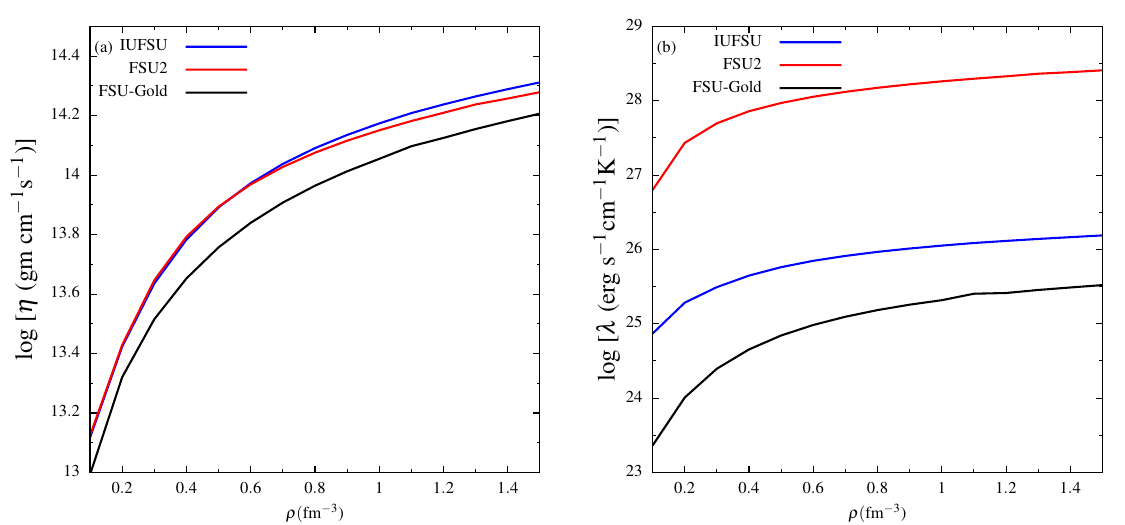}
		\caption{\it Variation of shear-viscosity (a) and thermal conductivity (b) with density for different hadronic models.}
		\label{fig:eta_lambda}
	\end{figure*}

	Next, we have calculated the average relaxation time using  Eq.~\eqref{eq:RTA:relaxationtime2}. 
	In this work, we have used parametrized nucleon-nucleon cross-sections \eqref{eq:NN_crosssection} and the leptonic scattering amplitudes are given in Eqs.~\eqref{eq:llcrosssection_1}, \eqref{eq:llcrosssection_2}, and \eqref{eq:llcrosssection_3}.
	In Fig.~\ref{fig:total_relax_time}, we have shown the variation of total relaxation time for the nucleons and the leptons. In Fig.~\ref{fig:total_relax_time}(a), we observe the relaxation time for neutrons and protons for the three parametrizations (hadronic models). We have seen that the total relaxation time decreases with density, and the neutron relaxation time is higher than the proton relaxation time. Similarly, in Fig.~\ref{fig:total_relax_time}(b), we have shown the total relaxation time for the electron and muon, and we have found that the electron relaxation time is higher than that of the muon relaxation time. 
	
	In Fig.~\ref{fig:eta_lambda}(a), we present the calculated shear viscosity at a temperature of $T = 0.1$ MeV. The shear viscosity $\eta$ is observed to increase with density. Even though the relaxation time decreases with increasing density, the shear viscosity still increases, primarily due to the greater number of particles available for the transport of momentum. The main contribution to the shear viscosity comes from neutrons, followed by electrons. This is mainly because neutrons are the most numerous, followed by protons, electrons, and muons. Although the density of protons is slightly greater than that of electrons for baryonic density $\rho$, electrons contribute more to the shear viscosity. The greater contribution of electrons compared to protons to the shear viscosity, $\eta$, is due to the higher mobility and longer relaxation time, $\tau_{e}$  of the electrons. If not for the $m^{2}_{a}$ factor in Eqn.~(\ref{eq:eta}),  electrons would have made the greatest contribution. 
	It is observed that the three hadron models have similar variations of $\eta$ with density.

	In Fig.~\ref{fig:eta_lambda}(b), we present the calculated thermal conductivity at a temperature of $T = 0.1$ MeV. 
	The thermal conductivity ($\lambda$) increases with density. The reason for the increase in the value of thermal conductivity ($\lambda$) with increasing density ($\rho$) of nuclear matter is the same as that for shear viscosity. As seen in Eqn.~(\ref{eq:lambda}), the conductivity mainly depends on the velocity $|p_a^{*}|/E_a^{*}$ of the particles and not on their momentum. Thus, the leptonic contribution to thermal conductivity is greater than the nucleonic contribution, since leptons, being lighter, have higher mobility (i.e., average velocity), with electrons being the lightest and contributing the most.
	As seen in Fig.~\ref{fig:eta_lambda}, the FSU-Gold and IUFSU models exhibit more or less similar variations, while the FSU2 model shows significantly larger and different thermal conductivity at higher baryonic densities. This is primarily because the electron chemical potential $\mu_{e}$ is much higher in the FSU2 model  as compared to the other models, resulting in a higher electron density. Since thermal conductivity is dominated by electrons, we obtain higher values for thermal conductivity $\lambda$ at higher densities ($\rho$).


	\subsection{Comparison with other works}\label{sec:Compariso} 
	\begin{figure*}[htp]
		\centering
		\includegraphics[width=1.0\textwidth]{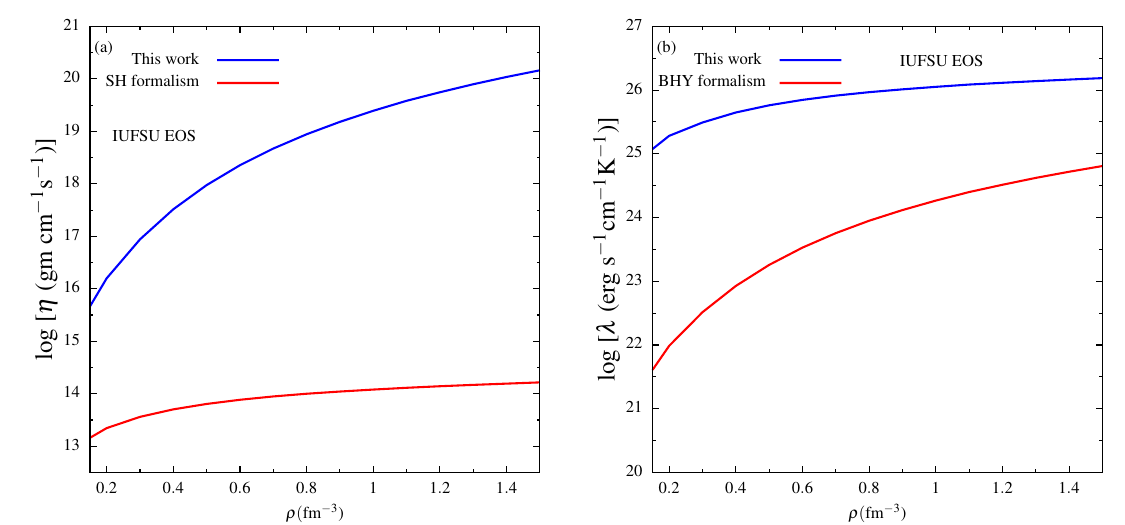}
		\caption{\it
			Comparison of the transport coefficients obtained in the present work with calculations based on established transport formalisms:
			(a) shear viscosity evaluated using the SH formalism ~\cite{Shternin:2008es}, and
			(b) thermal conductivity evaluated within the BHY formalism ~\cite{Baiko:2001cj}.
			Both sets of calculations are performed using the IUFSU equation of state.
		}

		\label{fig:compare}
	\end{figure*} 
	
	We compare our results with previous studies conducted at lower densities, particularly those of \cite{Wambach:1992ik} (hereafter WAP) and \cite{Sedrakian:1994uzz} (hereafter SE). WAP investigated pure neutron matter at densities around \( \rho_0 = 0.16 \,\text{fm}^{-3} \) using Landau theory. Their analysis included the calculation of momentum-dependent Landau parameters, quasiparticle transition amplitudes, and transport coefficients such as thermal conductivity, viscosity, and the spin diffusion coefficient. The estimated thermal conductivity is approximately  
	$\lambda_{\text{WAP}} \approx 2 \times 10^{21} \, \text{erg} \, \text{s}^{-1} \, \text{cm}^{-1} \, \text{K}^{-1},$
	assuming \( \rho_n = \rho_0 \) and \( T = 2 \) MeV.
	Sedrakian et al. \cite{Sedrakian:1994uzz} employed a thermodynamic \( T \)-matrix approach, yielding  
	$\lambda_{SE} \approx 2.6 \times 10^{20} \, \text{erg} \, \text{s}^{-1} \, \text{cm}^{-1} \, \text{K}^{-1}.$ 
	In \cite{Baiko:2001cj}, Baiko et al. (hereafter BHY)  studies the thermal conductivity of neutrons in neutron star cores. Their results show the $\lambda_{BHY}$ lies in the range of approximately  $ 10^{20}-10^{23} \text{erg} \, \text{s}^{-1} \, \text{cm}^{-1} \, \text{K}^{-1} $for temperature nearl 0.1 MeV. Our results show slighlty higher values, though its depends on the choice of the EOS, resulting in a variation of $10^{23}-10^{28}  \text{erg} \, \text{s}^{-1} \, \text{cm}^{-1} \, \text{K}^{-1}$ shown in Fig.~\ref{fig:eta_lambda}(b). The difference in the values also arises from the differing expressions for the relaxation time and the inclusion of relativistic corrections.

	For comparison of the shear viscosity, Shternin et al.~\cite{Shternin:2008es} (hereafter SH) studied the shear viscosity in neutron star cores. Their results indicate values of 
	$\eta_{\text{SH}} \sim 10^{17} - 10^{20}~\text{g}~\text{cm}^{-1}~\text{s}^{-1}.$
	In contrast, our results yield significantly lower values in the range
	$\eta \sim 10^{13} - 10^{15}~\text{g}~\text{cm}^{-1}~\text{s}^{-1}$
	as shown in Fig.~\ref{fig:eta_lambda}(a).

	To provide a consistent and direct comparison, Fig.~\ref{fig:compare} presents the transport coefficients calculated using the same IUFSU equation of state for both the established formalisms (SH for shear viscosity and BHY for thermal conductivity) and our present work. 
	Panel~(a) shows the shear viscosity, while panel~(b) shows the thermal conductivity. 
	The curves labeled ``This work'' correspond to our quasi-particle calculations, whereas the curves labeled ``SH formalism'' and ``BHY formalism'' represent the respective established formalisms evaluated using the IUFSU EOS. 
	Both panels indicate a monotonic increase of the transport coefficients with baryon density.
	In panel~(a), the shear viscosity from our calculations rises more steeply than that of the SH formalism. 
	In panel~(b), the thermal conductivity from our work is higher than the BHY formalism at lower densities, while the difference decreases slightly at higher densities.
	Overall, this comparison confirms that, with the same EOS, our calculations reproduce the qualitative trends of the established formalisms while highlighting quantitative differences arising from the specific treatment of particle interactions and relaxation times in the quasi-particle model.

	\begin{figure*}[htp]
		\centering
		\includegraphics[width=1.0\textwidth]{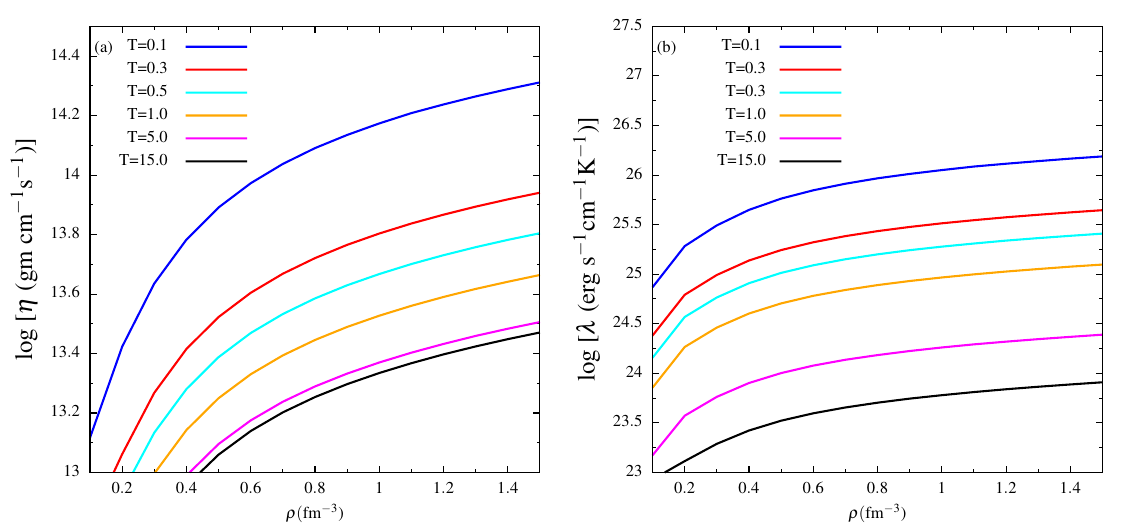}
		\caption{\it Variation of shear-viscosity (a) and thermal conductivity (b) with density for the different temperatures for IUFSU hadronic EOS. (temperature in the figure are in MeV unit)}
		\label{fig:eta_lambda_temp}
	\end{figure*}

	\subsection{Temperature effects on the transport coefficient}\label{sec:temp_effect} 
	
	In Fig.~\ref{fig:eta_lambda_temp}, we study the temperature dependence of shear viscosity and thermal conductivity in the neutron star core for the IUFSU hadronic equation of state. We find that both $\eta$ and $\lambda$ decrease with increasing temperature. An increase in temperature leads to more frequent collisions among the constituent particles, causing a reduction in their mobility and, consequently, the relaxation time. Since the relaxation times appear in the numerator of their analytical expressions, their values decrease with increasing temperature.

	We have also studied the fitting relations of $\eta$ and $\lambda$, which are expressed as follows:
	The values of the fitting coefficients are given in Table~\ref{tab:44}, where the temperature is in MeV and the baryon density is in $\text{fm}^{-3}$.

	\subsection*{Shear Viscosity $\eta(\rho,T)$}
	The logarithm of the shear viscosity follows the functional form:
	\begin{equation}
		\log \eta(\rho,T) = \sum_{n=1}^{5} \eta_{n}(T) f_n(\rho)
		\label{eq:eta_main}
	\end{equation}
	with density dependence:
	\begin{equation}
		\begin{aligned}
			f_1(\rho) &= \log \rho \\
			f_2(\rho) &= \rho^3 \\
			f_3(\rho) &= \rho^2 \\
			f_4(\rho) &= \rho \\
			f_5(\rho) &= 1
		\end{aligned}
		\label{eq:eta_rho}
	\end{equation}
	and temperature-dependent coefficients:
	\begin{equation}
		\eta_{i}(T) = a_i \log T + b_i T + c_i, \quad i=1,\ldots,5
		\label{eq:eta_coeff}
	\end{equation}
	The values of the density-dependent fitting coefficients for shear viscosity are provided in Table~\ref{tab:44}, with the temperature in MeV and the baryon density in $\text{fm}^{-3}$.
	The values the temperature dependent shear viscosity coefficient  is given in the Table.\ref{tab:2a}.

	\subsection*{Thermal Conductivity $\lambda(\rho,T)$}

	Similarly, the logarithm of the thermal conductivity follows:
	\begin{equation}
		\log \lambda(\rho,T) = \sum_{n=1}^{5} \lambda_{n}(T) g_n(\rho)
		\label{eq:lambda_main}
	\end{equation}
	with identical density dependence:
	\begin{equation}
		g_n(\rho) \equiv f_n(\rho), \quad n=1,\ldots,5
		\label{eq:lambda_rho}
	\end{equation}
	but with modified temperature dependence:
	\begin{equation}
		\lambda_{i}(T) = \alpha_i \log T + \beta_i T^2 + \gamma_i T + \delta_i, \quad i=1,\ldots,5
		\label{eq:lambda_coeff}
	\end{equation}
	The values of the density dependent fitting coefficients for the thermal conductivity are given in \ref{tab:3}, where temperate in MeV and baryon density in $\text{fm}^{-3}$
	The values the temperature dependent thermal conductivity coefficient is given in the Table.\ref{tab:3a}.

	\begin{table}[!ht]
		\caption{The fitting transport parameters for $\log \eta$ for the IUFSU EOS.}
		\setlength{\tabcolsep}{12 pt}
		\begin{tabular}{cccccc}
			\hline
			\hline
			T (MeV)&$\eta_1$ & $\eta_2$ & $\eta_3$ & $\eta_4$ & $\eta_5$ \\ 
			\hline
			
			0.1  &  0.375750 &  0.228451 & -0.767695 &  0.809142 & 13.905541 \\
			0.2  &  0.264373 &  0.324055 & -1.118487 &  1.324314 & 13.401727 \\
			0.3  &  0.186068 &  0.389667 & -1.360959 &  1.683266 & 13.094194 \\
			0.5  &  0.082289 &  0.476067 & -1.680501 &  2.157003 & 12.718144 \\
			1.0  & -0.047497 &  0.583103 & -2.076743 &  2.745641 & 12.279323 \\
			2.0  & -0.148457 &  0.665228 & -2.381209 &  3.199331 & 11.956961 \\
			3.0  & -0.196464 &  0.703645 & -2.524007 &  3.413100 & 11.812531 \\
			4.0  & -0.229592 &  0.729684 & -2.621200 &  3.559506 & 11.718681 \\
			5.0  & -0.256982 &  0.750719 & -2.700169 &  3.679460 & 11.645117 \\
			\hline
			\hline
		\end{tabular}
		\label{tab:44}
	\end{table}

	\begin{table}[!ht]
		\caption{Coefficients for $\eta_i(T) = a_i \log T + b_i T + c_i$ for the IUFSU EOS.}
		\setlength{\tabcolsep}{12 pt}
		\begin{tabular}{cccc}
			\hline
			$\eta_i$ &$a_{i}$ & $b_{i}$ & $c_{i}$  \\
			\hline
			$\eta_1$  & -0.194721 &  0.026337 & -0.064774 \\
			$\eta_2$  &  0.163915 & -0.024757 &  0.600747 \\
			$\eta_3$  & -0.604640 &  0.089792 & -2.139794 \\
			$\eta_4$ &  0.893403 & -0.129016 &  2.834251 \\
			$\eta_5$  & -0.764443 &  0.168707 & 12.125392 \\
			\hline
			\hline
		\end{tabular}
		\label{tab:2a}
	\end{table}

	\begin{table}[!ht]
		\caption{The fitting transport parameters for $\log \lambda$ for the IUFSU EOS.}
		\setlength{\tabcolsep}{12 pt}
		\begin{tabular}{cccccc}
			\hline
			\hline
			T (MeV)&$\lambda_1$ & $\lambda_2$ & $\lambda_3$ & $\lambda_4$ & $\lambda_5$ \\ 
			\hline
			
			0.1  &  0.630724 &  0.020953 & -0.010815 & -0.311779 & 26.353194 \\
			0.2  &  0.650725 & -0.016177 &  0.135931 & -0.524848 & 26.110480 \\
			0.3  &  0.658765 & -0.031137 &  0.193492 & -0.607132 & 25.958109 \\
			0.5  &  0.665849 & -0.044027 &  0.243074 & -0.678138 & 25.757431 \\
			1.0  &  0.671501 & -0.054675 &  0.283793 & -0.736047 & 25.474237 \\
			2.0  &  0.671959 & -0.059302 &  0.301047 & -0.758774 & 25.178427 \\
			3.0  &  0.666516 & -0.057307 &  0.293074 & -0.744316 & 24.992352 \\
			4.0  &  0.657351 & -0.051953 &  0.272296 & -0.710529 & 24.848361 \\
			5.0  &  0.645768 & -0.044761 &  0.244318 & -0.665396 & 24.726708 \\
			\hline
			\hline
		\end{tabular}
		\label{tab:3}
	\end{table}

	\begin{table}[!ht]
		\caption{Coefficients for $\lambda_i(T) = \alpha_i \log T + \beta_i T^2 + \gamma_i T + \delta_i$ for the IUFSU EOS.}
		\setlength{\tabcolsep}{12 pt}
		\begin{tabular}{ccccc}
			\hline
			$\lambda_i$ &$\alpha_{i}$ & $\beta_{i}$ & $\gamma_{i}$ & $\delta_{i}$  \\
			\hline
			$\lambda_1$  &  0.025432 &  0.000541 & -0.020276 &  0.693514 \\
			$\lambda_2$  & -0.046891 & -0.002819 &  0.038923 & -0.095203 \\
			$\lambda_3$  &  0.183398 &  0.011311 & -0.154183 &  0.444377 \\
			$\lambda_4$  & -0.264907 & -0.015982 &  0.224009 & -0.969877 \\
			$\lambda_5$  & -0.360180 &  0.002434 & -0.057598 & 25.535980 \\
			\hline
		\end{tabular}
		\label{tab:3a}
	\end{table}

	\section{SUMMARY AND CONCLUSION}\label{sec:conclusion}

	We  have utilized a relativistic kinetic theory framework to compute the transport coefficients of neutron star matter. The BUU transport equation, which  describes the evolution of neutron star matter, was solved using the relaxation time approximation. For calculating the relaxation time, we have employed a parametrized cross-section for nucleons and a field-theoretic approach for leptons. In this work, we analyzed the shear viscosity and thermal conductivity inside the neutron star core using the RMF model. Our results are as follows: relaxation times for neutrons are higher than for protons, while electrons have higher relaxation times than muons. Shear viscosity $\eta$ and thermal conductivity $\lambda$ increase with density, with $\eta_n > \eta_e > \eta_\mu > \eta_p$ and $\lambda_e > \lambda_\mu > \lambda_p > \lambda_n$. The FSU2 model shows higher  values of $\lambda$ due to higher electron chemical potential $\mu_e$ compared to the FSU-Gold and IUFSU models. We have compared our results with previous works on this subject. We found that the transport coefficients calculated using our method follow similar trends to those predicted by previous studies, although their absolute values differ significantly. We have also introduced parametrized formulas for the transport coefficients, calculated analytically and valid for the temperature range of $0.1 ~ \text{MeV}$ to $5.0~\text{MeV}$ and for baryon density range $0.1~\text{fm}^{-3}$ to $1.5~\text{fm}^{-3}$.

	\section{Acknowledgment}
	
	U.G. would like to express gratitude to the Department of Atomic Energy, India, and VECC for their financial support. U.G. and S.P. are grateful to Dr. Surasree Mazumder (IISER Berhampur) for her invaluable help and essential inputs regarding various aspects of field theory.

	\appendix

	\bibliographystyle{elsarticle-num}
	\bibliography{transport}
\end{document}